\documentstyle[aps,prl,preprint,floats,epsfig]{revtex}

\begin{document}

\preprint{\tighten\vbox{\hbox{\hfil CLNS 02-1776}
                        \hbox{\hfil CLEO 02-02}
}}

\title{Measurement of the $D^+ \rightarrow \bar{K}^{*0} l^+ \nu_l$
Branching Fraction}  

\author{CLEO Collaboration}
\date{March 12, 2002}

\maketitle
\tighten

\begin{abstract} 
Using $13.53$ $\rm fb^{-1}$ of CLEO data, we have measured 
the ratios of the branching fractions 
$R^+_e = \frac {{\cal B}(D^+ \rightarrow \bar{K}^{*0} e^+ \nu_e)} 
{{\cal B}(D^+ \rightarrow K^- \pi^+ \pi^+)}$, 
$R^+_{\mu} = \frac {{\cal B}(D^+ \rightarrow \bar{K}^{*0} \mu^+ \nu_{\mu})} 
{{\cal B}(D^+ \rightarrow K^- \pi^+ \pi^+)}$ 
and the combined branching fraction ratio 
$R^+_l = \frac {{\cal B}(D^+ \rightarrow \bar{K}^{*0} l^+ \nu_l)} 
{{\cal B}(D^+ \rightarrow K^- \pi^+ \pi^+)}$. We find 
$R^+_e$ $=$ $0.74 \pm 0.04 \pm 0.05$, 
$R^+_{\mu}$ $=$ $0.72 \pm 0.10 \pm 0.06$ and 
$R^+_l$ $=$ $0.74 \pm 0.04 \pm 0.05$, where the first 
and second errors are statistical and systematic, respectively. 
The known branching fraction 
${\cal B}(D^+ \rightarrow K^- \pi^+ \pi^+)$ leads to:
${\cal B}(D^+ \rightarrow \bar{K}^{*0} e^+ \nu_e)$ $=$ 
$(6.7 \pm 0.4 \pm 0.5 \pm 0.4 ) \%$, 
${\cal B}(D^+ \rightarrow \bar{K}^{*0} \mu^+ \nu_{\mu})$ $=$ 
$(6.5 \pm 0.9 \pm 0.5 \pm 0.4 ) \%$ and 
${\cal B}(D^+ \rightarrow \bar{K}^{*0} l^+ \nu_l)$ $=$ 
$(6.7 \pm 0.4 \pm 0.5 \pm 0.4 ) \%$, 
where the third error is due to the uncertainty 
in ${\cal B}(D^+ \rightarrow K^- \pi^+ \pi^+)$.
\end{abstract}
\newpage

{
\renewcommand{\thefootnote}{\fnsymbol{footnote}}

\begin{center}
G.~Brandenburg,$^{1}$ A.~Ershov,$^{1}$ D.~Y.-J.~Kim,$^{1}$
R.~Wilson,$^{1}$
K.~Benslama,$^{2}$ B.~I.~Eisenstein,$^{2}$ J.~Ernst,$^{2}$
G.~D.~Gollin,$^{2}$ R.~M.~Hans,$^{2}$ I.~Karliner,$^{2}$
% ...
\end{center}
 
\small
\begin{center}
$^{1}${Harvard University, Cambridge, Massachusetts 02138}\\
$^{2}${University of Illinois, Urbana-Champaign, Illinois 61801}\\
% ...
\end{center}

\setcounter{footnote}{0}
}
\newpage

The transition amplitude of the semileptonic decay 
$D^+ \rightarrow \bar{K}^{*0} l^+ \nu_l$ is proportional to 
the product of the leptonic and hadronic currents. 
The hadronic current is represented by three analytic functions 
called form factors, $A_1(q^2)$, $A_2(q^2)$, and $V(q^2)$~\cite{FF}, 
where $q^2$ is the invariant mass squared of the virtual $W$ boson.
The form factors cannot be easily computed 
in quantum chromodynamics (QCD) since they are affected by 
significant non-perturbative contributions. 
Further, unlike the case of $B \rightarrow D^* l \nu$, 
Heavy Quark Effective Theory (HQET)~\cite{HQET1,HQET2,HQET3} 
cannot be applied in a straightforward way since the transition is 
to the light strange quark. 
Precise experimental measurements are needed to  
guide theoretical progress in this area.
Measurements of these form factors are also valuable since,  
using HQET, they can 
be related to the form factors for the important decay 
$b \rightarrow u l \nu$, a heavy to light transition like the 
$c \rightarrow s l \nu$ process we study here. This approach could 
reduce theoretical uncertainties on the 
value of $|V_{ub}|$ measured in $b \rightarrow u l \nu$.

In this work, the ratio of the branching fraction of 
$D^+ \rightarrow \bar{K}^{*0} l^+ \nu_l$ to that of 
$D^+ \rightarrow K^- \pi^+ \pi^+$ ($R_l^+$) was measured, where 
$l$ stands for either an electron or a muon. In other 
experiments~\cite{r2_rv_all}, 
the form factor ratios $r_2 = \frac {A_2}{A_1}$ and $r_V = \frac {V}{A_1}$ 
of $D^+ \rightarrow \bar{K}^{*0} l^+ \nu_l$ were obtained from angular 
correlations between the daughter particles. Combining these, we can 
calculate the form factors $A_1$, $A_2$ and $V$.

The data used in this study was collected with the two configurations of 
the CLEO detector~\cite{CLEO} at the Cornell Electron Storage Ring (CESR). 
It consists of $9.13$ $\rm fb^{-1}$ of integrated luminosity on the 
$\Upsilon(4S)$ resonance and $4.40$ $\rm fb^{-1}$ below $B \bar{B}$ threshold.
The investigated decay chain for this analysis was 
$D^{*+} \rightarrow D^+ \pi^0$, $D^+ \rightarrow \bar{K}^{*0} l^+ \nu$, 
and $\bar{K}^{*0} \rightarrow K^- \pi^+$. 
The use of $D^+$ produced in  $D^{*+} \rightarrow D^+ \pi^0$ decays reduced 
the background. This analysis assumes that the thrust axis~\cite{thrust} 
of the event approximates the direction of the $D$ meson, as will be 
explained below. This approach does not work well for isotropic events.
Thus, events whose ratio of the second and zeroth 
Fox-Wolfram moments~\cite{Fox-Wolfram} was less 
than or equal to $0.2$ were rejected. 
Specific ionization in the drift chamber and shower 
information in the electromagnetic calorimeter were 
used to select good electron candidates.  Muon 
candidates were required to penetrate at least 5 
interaction lengths in our muon detector.
To select good $\pi^0$'s, all $\gamma \gamma$ pairs for which 
$|M_{\gamma \gamma} - M_{\pi^0}|$ $<$ $2.5 \sigma$ were accepted, where 
$\sigma$ is the standard deviation of the $\pi^0$ mass measurement, 
obtained as a function of $\pi^0$ momentum using clean $\pi^0$ data samples.
Other kinematic criteria chosen to optimize our sensitivity
are given in Table~\ref{tab:anal_cuts}. 
\begin{table}
\begin{center}
\caption{Requirements on the kinematic variables. Here, $\theta$ 
         is the polar angle between the $e^+ e^-$ axis and the momentum of the 
         particle candidate.}
\label{tab:anal_cuts}
\begin{tabular}{c l} 
Variables		& \multicolumn{1}{c}{Requirements} \\
\hline
$|\vec{p}_{\pi^+}|$	&$>$ $0.5$~GeV/$c$ \\
$|\vec{p}_{K^-}|$	&$>$ $0.5$~GeV/$c$ \\
$|\vec{p}_{\pi^0}|$	&$>$ $0.18$~GeV/$c$ \\
$|\vec{p}_e|$		&$>$ $0.7$~GeV/$c$ for $|{\rm cos} \theta| \le 0.81$\\
$|\vec{p}_{\mu}|$	&$>$ $1.4$~GeV/$c$ for $|{\rm cos} \theta| \le 0.61$\\
$|\vec{p}_{\mu}|$	&$>$ $1.9$~GeV/$c$ for $0.61 < |{\rm cos} \theta| \le 0.81$\\
$|\vec{p}_{K \pi l}|$	&$>$ $2.0$~GeV/$c$ \\     
$m_{K \pi l}$		&$1.2$ $-$ $1.8$~GeV/$c^2$ \\ 
$|\vec{p}_{K \pi}|$	&$>$ $0.7$~GeV/$c$ \\
\end{tabular}
\end{center}
\end{table}

To reconstruct the momentum of the $\nu$, 
two methods were used to obtain up to three values of $\vec{p}_{\nu}$. 
In the first method, one or two values for the $\nu$ momentum were 
obtained assuming that the thrust direction of the event represents 
the $D$ direction. Given $\vec{p}_{K \pi l}$, the constraint that 
$m_{K \pi l \nu} = m_{D^+}$ 
provided an ellipsoid of allowed $D$ momenta.  We then took the 
two intersections between the ellipsoid and the direction of the $D$. 
When there was no intersection, the point on the ellipsoid that
lay closest to the $D$ direction was used. The values for the $\nu$ momentum 
were then the difference between these $D$ momenta and $\vec{p}_{K \pi l}$.
The second method used the missing momentum of each event as an estimate 
of the $\nu$ momentum. If the missing momentum gave a value of 
$m_{K \pi l \nu}$ much greater than the mass of the $D^+$ 
(we used $m_{D^{*+}}$ 
for the limit for convenience), then the $\nu$ momentum estimate from the 
second method was discarded. Among these three $\nu$ momentum estimates, 
the one that gave the value of 
$\delta m = m_{K \pi l \nu \pi^0} - m_{K \pi l \nu}$ closest 
to the known value of  $m_{D^{*+}}-m_{D^+}$~\cite{pdb} was chosen.
According to our simulation, this selection method gave the 
$\nu$ momentum closest to the generated $\nu$ momentum 75\% of the time. 

The previous CLEO analysis of this decay~\cite{bean} superseded  
by this study did not use $\nu$ 
reconstruction.  Instead it used the pseudo mass difference, $\delta m_p = 
m_{K \pi l \pi^0} - m_{K \pi l}$, and the resolution of $\delta m_p$ 
was much inferior to $\delta m$.
The $\nu$ reconstruction reduced the statistical uncertainty in the 
${\cal B}(D^+ \rightarrow \bar{K}^{*0} l^+ \nu_l)$ measurement 
by 20\% based on studies using signal and generic continuum 
Monte Carlo (MC) events. However, since our method for choosing among 
the three solutions for $|\vec{p}_{\nu}|$ tends to bias the $\delta m$ 
distribution of the background toward the signal region, it is important 
to understand the contribution of the background that peaks in the signal 
region. The uncertainty coming from this effect was included in the 
systematic errors.

\begin{figure} 
  \centerline{\epsfig{width=8.6cm, figure=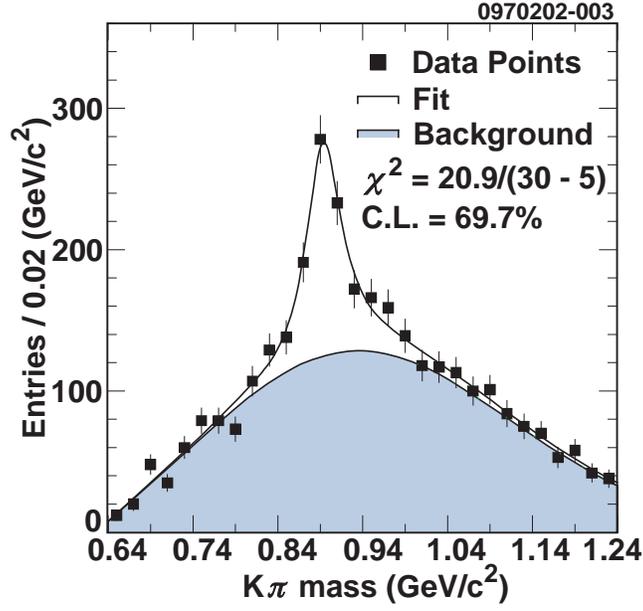}}
  \caption{$K \pi$ mass distribution in the third $\delta m$ bin 
           of the $D^+ \rightarrow \bar{K}^{*0} e^+ \nu_e$ analysis 
           using $8.78$~${\rm fb^{-1}}$ data.}
  \label{fig:mksfit_e_25_sig}
\end{figure}

Two quantities were fitted to obtain the number of 
signal events: $m_{K \pi}$ and $\delta m$. 
First, the data were divided into 50 bins (25 bins for the smaller 
muon data) over the $\delta m$ range 
from $0.135$ GeV/$c^2$ to $0.235$ GeV/$c^2$.
For each $\delta m$ bin, the $K \pi$ mass distribution was plotted and the 
number of $K^*$'s was extracted by a fit. 
Second, these yields were plotted as a function of $\delta m$, and 
the number of signal events was obtained by another fit.
In the $K \pi$ mass fit, the signal shape was described by 
a $p$-wave $K^*$ Breit-Wigner distribution. The mass and width 
of $K^*$ obtained by the fit were consistent with the Particle Data 
Group (PDG) values~\cite{pdb}. An analytic threshold function 
\begin{equation}
\small
(m_{K \pi} - 0.64)^\alpha \times e^{\beta  (m_{K \pi} - 0.64) +
                                    \gamma (m_{K \pi} - 0.64)^2}
\label{EQ:1}
\normalsize
\end{equation} 
parameterized the shape of the background. The appropriateness of this 
function was tested using generic continuum MC events after discarding 
the signal among those events.
The parameters $\alpha$, $\beta$ and $\gamma$ in this 
background function were unconstrained in the data fits. 
A typical $K \pi$ mass fit in the $\delta m$ signal 
region is shown in Fig.~\ref{fig:mksfit_e_25_sig}.

\begin{figure} 
  \centerline{\epsfig{width=8.6cm, figure=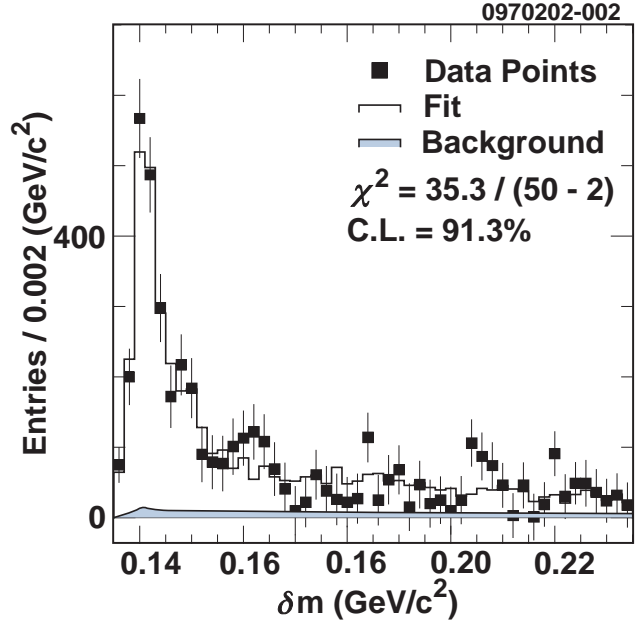}}
  \caption{The $\delta m$ distribution for 
           $D^+ \rightarrow \bar{K}^{*0} e^+ \nu_e$ using 
           $8.78$~${\rm fb^{-1}}$ data.}
  \label{fig:ksenu_dmfit_25}
\end{figure}

The signal shapes for the $\delta m$ fits were obtained from 
signal MC samples, and the background shape was described by 
a continuous function which was designed to accommodate the 
excess at $m_{D^{*+}} - m_{D^+}$ due to our neutrino momentum
selection method. 
The shapes of these functions in the $\delta m$ fits for data 
were determined using a generic continuum MC sample. 
The $\delta m$ fit of the electron data 
is shown in Fig.~\ref{fig:ksenu_dmfit_25}. 

The decay chain, $D^{*+} \rightarrow D^+ \pi^0$, 
$D^+ \rightarrow K^- \pi^+ \pi^+$, was used for the normalization mode.
One advantage of this mode is that many systematic errors in 
its ``measurement'' are common with the 
$D^+ \rightarrow \bar{K}^{*0} l^+ \nu$ mode and cancel in their ratio. 
Furthermore, the branching fraction of $D^+ \rightarrow K^- \pi^+ \pi^+$ 
is well measured: $(9.0 \pm 0.6)\%$~\cite{pdb}.
Most of the kinematic requirements were the same as those in the 
semileptonic analysis, but the requirement on $|\vec{p}_{D^*}|$ was tuned 
to minimize our sensitivity to the uncertainty in the fragmentation 
of the charm quarks.

For the normalization mode analysis, two successive fits of $m_{D^+}$ 
and $\delta m$ were used.
The data were divided into 50 $\delta m$ bins from 
$0.135$ GeV/$c^2$ to $0.185$ GeV/$c^2$. For each $\delta m$ bin, 
a $D^+$ mass fit was used to extract the number of $D^+$'s. 
These yields of $D^+$'s were plotted as a function of $\delta m$. 
Then, this $\delta m$ plot was fitted to obtain the number of the 
normalization mode events. 

Table~\ref{tab:sys_error_comb} lists the important contributions to 
the systematic error. The dominant sources are the $\delta m$ signal shape 
estimation due to finite MC statistics, potential flaws in the 
estimate of the background shape obtained from the simulation of 
$e^+ e^- \rightarrow q \bar{q}$ and background arising from 
$D^+ \rightarrow \bar{K}^{*0} \pi^0 l^+ \nu_{l}$.

The systematic error due to a potential difference 
in the background between the 
MC and data was estimated using a $\pi^0$ mass sideband, 
$3.5 \sigma$ $<$ $|M_{\gamma \gamma} - M_{\pi^0}|$ $<$ $15.0 \sigma$. 
The $\pi^0$ mass sideband data were analyzed as $\pi^0$ signal band 
data and the difference in yields of data and generic continuum MC sample 
was used in the estimation of the systematic error due to the 
deficiency of the background MC in reproducing the data.

The upper limit of the branching fraction for 
$D^+ \rightarrow \bar{K}^{*0} \pi^0 \mu^+ \nu_{\mu}$ with respect to 
$D^+ \rightarrow K^- \pi^+ \mu^+ \nu_{\mu}$ is 0.042~\cite{e687} at 
the $90 \%$ confidence level. 
We used the product of this upper limit and the ratio of ``efficiencies'' 
that $K^* \pi^0 l \nu$ and $K^* l \nu$ are detected as signal in our analysis 
as the systematic error due to this source.

The systematic errors from each source 
were estimated individually for the electron and muon modes and then 
combined to obtain the overall lepton mode systematic errors. 
\begin{table}
\squeezetable
\begin{center}
\caption{The electron, muon and electron-muon combined systematic errors 
         in units of
         $10^{-2}$. The combined systematic errors of the electron and 
         muon modes are denoted as the $l$. }
\label{tab:sys_error_comb}
\begin{tabular}{cllll} 
\multicolumn{2}{c}{Source}   & \multicolumn{1}{c}{$e$}      & 
\multicolumn{1}{c}{$\mu$}    & \multicolumn{1}{c}{$l$}    \\
\hline
fits &                       &          &            &        \\ 
     & $\delta m$ sig. shape & $2.2$    & $3.3$      & $1.9$  \\  
     & $\delta m$ bgr. shape & $1.2$    & $1.4$      & $1.0$  \\
     & data MC diff.(sig.)   & $0.34$   & $3.8$      & $0.68$ \\ 
     & data MC diff.(bgr.)   & $2.6$    & $2.5$      & $2.6$  \\ 
\hline
lepton id              &     & $1.1$    & $0.27$     & $0.89$ \\
\hline
fake leptons	       &     & $0.064$  & $0.95$     & $0.17$ \\
\hline
sig MC model dep.      &     & $0.74$   & $0.72$     & $0.74$ \\
\hline
efficiency by fragmnt. &     & $0.53$   & $0.54$     & $0.53$ \\
\hline
feed down	       &     & $0.18$   & $0.17$     & $0.18$ \\
\hline
$D^+ \rightarrow \bar{K}^{*0} \pi^0 l^+ \nu_{l}$ &
			     & $2.6$    & $2.5$      & $2.6$  \\
\hline
normalization mode     &     &         &            &         \\
     & sig. shape            & $1.1$   & $1.1$      & $1.1$   \\  
     & bgr. shape            & $0.81$  & $0.79$     & $0.81$  \\  
\hline\hline
TOTAL                  &     & $4.8$   & $6.0$      & $4.7$   \\  
\end{tabular}
\end{center}
\end{table}

The ratio $R^+_l$ was calculated using 
\begin{equation}
  R^+_l = \frac{N_{sl}}{\epsilon_{sl} \cdot {\cal B}(\bar{K}^{*0} 
          \rightarrow K^- \pi^+)} \cdot \frac{\epsilon_{had}}{N_{had}},
\label{EQ:4}
\end{equation}
where $N_{sl}$ and $\epsilon_{sl}$ are the observed number of events and the 
efficiency for $D^+ \rightarrow \bar{K}^{*0} l^+ \nu_l$, respectively. 
Similarly, $N_{had}$ and $\epsilon_{had}$ are the corresponding quantities for 
$D^+ \rightarrow K^- \pi^+ \pi^+$. 
The measured branching ratios $R_e^+$, $R_{\mu}^+$ and $R_l^+$, 
including their errors, are summarized in Table~\ref{tab:R_reslts_fnl}. 
The values of $R_e^+$ and $R_{\mu}^+$ agree well. The PDG value of 
${\cal B}(D^+ \rightarrow K^- \pi^+ \pi^+) = (9.0 \pm 0.6) \%$~\cite{pdb} 
was used to calculate 
the branching fractions of $D^+ \rightarrow \bar{K}^{*0} e^+ \nu_e$ 
(${\cal B_{\it e}}$), 
$D^+ \rightarrow \bar{K}^{*0} \mu^+ \nu_{\mu}$ (${\cal B_{\mu}}$), and 
$D^+ \rightarrow \bar{K}^{*0} l^+ \nu_l$ (${\cal B_{\it l}}$). 
\begin{table}
\begin{center}
\caption{Results of the $R^+_e$, $R^+_{\mu}$ and $R^+_l$ measurements 
         and the branching fractions ${\cal B_{\it e}}$, ${\cal B_{\mu}}$ 
         and ${\cal B_{\it l}}$ (see text). The first and second errors 
         in the $R$ and ${\cal B}$ measurements are due to statistical 
         and systematic errors, respectively. The third errors in the 
         ${\cal B}$ measurements are due 
         to the uncertainty in ${\cal B}(D^+ \rightarrow K^- \pi^+ \pi^+)$.}
\label{tab:R_reslts_fnl}
\begin{tabular}{cc} 
$R^+_e$     & $0.74 \pm 0.04 \pm 0.05 $ \\
$R^+_{\mu}$ & $0.72 \pm 0.10 \pm 0.06 $ \\
$R^+_l$     & $0.74 \pm 0.04 \pm 0.05 $ \\
\hline
${\cal B_{\it e}}$     & $(6.7 \pm 0.4 \pm 0.5 \pm 0.4)\%$ \\
${\cal B_{\mu}}$       & $(6.5 \pm 0.9 \pm 0.5 \pm 0.4)\%$ \\
${\cal B_{\it l}}$     & $(6.7 \pm 0.4 \pm 0.5 \pm 0.4)\%$ \\
\end{tabular}
\end{center}
\end{table}

Table~\ref{tab:r_l_comp} compares the results of this work with  
previous measurements~\cite{bean,e687,omeg,argus,e691,e653}
and the PDG averages~\cite{pdb}. Note that the E691
measurement of $R_e^+$ and ours are the two most significant measurements,
and they differ by about three standard deviations. We studied whether 
this difference (or part thereof) may arise from differences between the 
$D^+ \rightarrow \bar{K}^{*0} e^+ \nu_e$ model used here (ISGW2)~\cite{isgw2} 
and the one used in the E691 analysis (WSB)~\cite{wsb}. Since
the lepton momentum distribution expected from this decay has a significant
effect on its detection efficiency in both experiments, we compared the 
electron energy and $q^2$ distributions from the two models, 
but found no significant differences. We did not identify any other 
explanations for the discrepancy. 
\begin{table} 
\squeezetable
\begin{center}
\caption{Comparison of the measured values of $R^+_e$ and $R_{\mu}^+$.}
\label{tab:r_l_comp}
\begin{tabular}{cl|cl} 
\multicolumn{1}{c}{Group}  &  \multicolumn{1}{c|}{$R^+_e$}  & 
\multicolumn{1}{c}{Group}  &  \multicolumn{1}{c}{$R_{\mu}^+$}  \\
\hline
CLEO(2001)        & $0.74 \pm 0.04 \pm 0.05 $  & 
CLEO(2001)        & $0.72 \pm 0.10 \pm 0.06 $  \\ 
PDG~\cite{pdb}     & $0.54 \pm 0.05$            & 
PDG~\cite{pdb}     & $0.53 \pm 0.06$            \\
CLEO~\cite{bean}   & $0.67 \pm 0.09 \pm 0.07$   &
E687~\cite{e687}   & $0.56 \pm 0.04 \pm 0.06$   \\
OMEG~\cite{omeg}   & $0.62 \pm 0.15 \pm 0.09$   &
E653~\cite{e653}   & $0.46 \pm 0.07 \pm 0.08$   \\
ARGUS~\cite{argus} & $0.55 \pm 0.08 \pm 0.10$ & & \\
E691~\cite{e691}   & $0.49 \pm 0.04 \pm 0.05$ & & \\
\end{tabular}
\end{center}
\end{table}

Table~\ref{tab:ksenu_decayrate} shows the decay rate for 
$D^+ \rightarrow \bar{K}^{*0} e^+ \nu_e$ measured in this work 
and the theoretical predictions~\cite{isgw2,ukqcd2001,ape,ukqcd,elc,lmms}. 
Most predictions are consistent with both our result and that of E691. 
\begin{table} 
\squeezetable
\begin{center}
\caption{Comparison of the decay rate for 
         $D^+ \rightarrow \bar{K}^{*0} e^+ \nu_e$ measured in this work 
         with those theoretically predicted and the present PDG value. All 
         values are in units of $10^{10} |V_{cs}|^2$ ${\rm sec^{-1}}$.}
\label{tab:ksenu_decayrate}
\begin{tabular}{cl} 
\multicolumn{1}{c}{Group}     & 
\multicolumn{1}{c}{Decay Rate ($ 10^{10} |V_{cs}|^2$ ${\rm s^{-1}}$)}   \\
\hline
CLEO(2001)        & $6.67 \pm 0.75$ \\ 
PDG~\cite{pdb}     & $4.79 \pm 0.40$ \\
UKQCD (2001)~\cite{ukqcd2001} & $5.8 \pm 0.5$ \\
APE~\cite{ape}     & $6.9  \pm 1.8$  \\
UKQCD (1994)~\cite{ukqcd} & $6.0^{+ 0.8}_{- 1.6}$ \\
ELC~\cite{elc}     & $6.4  \pm 2.8$  \\
LMMS~\cite{lmms}   & $5.0  \pm 0.9$  \\
ISGW2~\cite{isgw2} & $5.7$ \\
\end{tabular}
\end{center}
\end{table}
Using our branching ratio 
measurements and the ratios between the form factors measured by
E791~\cite{r2_rv}, we calculate 
$A_1(0) = 0.69 \pm 0.07$, $A_2(0) = 0.48 \pm 0.08$, 
and $V(0) = 1.25 \pm 0.15$, where the errors are the quadratic sums of 
the statistical and systematic errors in this analysis and in the E791 
measurement.

The ratio of the vector and pseudo-scalar decay widths,  
$r_{vp} = \frac{\Gamma(D \rightarrow K^* e^+ \nu_e)}
{\Gamma(D \rightarrow K   e^+ \nu_e)}$, tests quark models. 
This ratio was predicted to be in the range $0.9$ to $1.2$ 
by early quark models~\cite{wsb,Altomari,Korner,isgw,Gilman} 
and lattice gauge calculations~\cite{Bernard,Lubicz}. The E691 result 
implied that $r_{vp} = 0.48 \pm 0.10$.  The newer ISGW2 model~\cite{isgw2} 
accommodated this measurement well, predicting $r_{vp} = 0.54$.  
Our measurement of $R_l^+$, on the other hand, implies that 
$r_{vp} = 0.99 \pm 0.06 \pm 0.07 \pm 0.06$, where the first, second 
and third errors are due to our statistical and systematic errors and the 
uncertainty in  ${\cal B}(D^+ \rightarrow \bar{K}^{0} e^+ \nu_e)$, 
respectively, using the PDG value of 
${\cal B}(D^+ \rightarrow \bar{K}^{0} e^+ \nu_e)$ $=$ 
$(6.8 \pm 0.8)\%$~\cite{pdb}.
Our ratio is consistent with older models and disagrees with the ISGW2 
prediction by four standard deviations.

We gratefully acknowledge the effort of the CESR staff 
in providing us with excellent luminosity and running 
conditions. This work was supported by the National 
Science Foundation, the U.S. Department of Energy, the 
Research Corporation, the Natural Sciences and Engineering 
Research Council of Canada, and the Texas Advanced 
Research Program. 


\end{document}